\documentclass[twocolumn,a4paper,amsmath,amssymb,showpacs,prb,superscriptaddress]{revtex4}
\usepackage{amsmath}
\usepackage{bm}
\usepackage{graphicx}
\usepackage{epstopdf}
\usepackage{color}
\usepackage[normalem]{ulem}

\begin{document}

\title{Multipole order and global/site symmetry in the hidden order
phase of URu$_2$Si$_2$}

\author{Michi-To Suzuki}
\affiliation{RIKEN Center for Emergent Matter Science (CEMS), Wako,
Saitama 351-0198, Japan}
\affiliation{CCSE, Japan Atomic Energy Agency, 5-1-5 Kashiwanoha,
Kashiwa, Chiba 277-8587, Japan}
\author{Hiroaki Ikeda}
\affiliation{Department of Physics, Ritsumeikan University, Kusatsu 525-8577, Japan}

\date{\today}

\begin{abstract}
On the basis of group theory and the first-principles calculations, we
 investigate high-rank multipole orderings in URu$_2$Si$_2$, which have
 been proposed as a genuine primary order parameter in the hidden order phase
 below $17.5$K. We apply Shubnikov group theory to the multipole ordered
 states characterized by the wave vector ${\bm Q}_0=(0,0,1)$ and
 specify the global/site symmetry and
 the secondary order parameters, such as induced dipole moments and
 change in charge distribution. 
 We find that such antiferroic {\it magnetic} multipole orderings have
 particularly advantageous to conceal the primary order parameter due to
 preserving high symmetry in charge distribution.
 Experimental observations of the induced low-rank multipoles, which
 are explicitly classified in this paper, will be key pieces to
 understand the puzzling hidden order phase.
\end{abstract}


\maketitle

\section{INTRODUCTION}
%
%
 Recently, $f$ electron compounds have drawn considerable interest,
 stimulated by observations of exotic ordered ground states forming at
 low temperatures.
The richness of $f$ electron physics can be attributed to their multiple
degrees of freedom in the presence of strong spin-orbit
 interaction as well as strong electron-electron Coulomb interactions. 
A striking example is formation of multipole order
 (MPO), in which spin-orbital space of $f$ electrons
 has main role to establish the physical states with strongly correlated electrons.
 While recent findings of MPO in the $f$-electron compounds promise to
 exotic materials with unprecedented functions, the complex constitution
 of the microscopic quantum states makes it difficult to identify the
 ordered states experimentally.
 A well known example of the MPO formation is the ground state of NpO$_2$. 
While the order parameter had been enigma for fifty years,~\cite{santini09, kuramoto09} it has been recently clarified that the ordering is characterized by the 3${\bf q}$ antiferroic (AF) structure of T$_2$ multipoles,~\cite{paixao02, tokunaga05, tokunaga06} which is a high-rank multipole without dipolar magnetic moment.
Another example is the formation of the hidden order (HO) phase in
 URu$_2$Si$_2$, where various type of the high-rank multipoles have been
 suggested as promising primary order parameters by theoretical
 studies.~\cite{santini94, ohkawa99, fazekas05, kiss05, hanzawa05,
 cricchio09, haule09, harima10, thalmeier11, kusunose11, ikeda12,
 rau12}

 The HO state accompanied by the second order phase transition at
 $T_{\rm HO}$=17.5K in URu$_2$Si$_2$ has been an object of study over 30
 years since its discovery. ~\cite{palstra85, maple86, schlabitz86} 
The phase transition is associated with abrupt change in the bulk
properties at T$_{\rm HO}$ and is accompanied by large reduction of the
carrier number.~\cite{behnia05, kasahara07}
 The tiny magnetic moments experimentally observed in the HO phase are
 currently considered to be an extrinsic effect from inhomogeneity of
 the crystal,~\cite{broholm87, mason90, isaacs90, walker93, amitsuka99, matsuda01,
 amitsuka07} and there is little consensus about the order parameter
 which characterizes the HO phase.~\cite{mydosh11} 
The conventional AF magnetic phase appear accompanied by
the first order phase transition at P$_c$ $\sim$1.5GPa.~\cite{amitsuka03,
amitsuka99, hassinger08}
Inelastic neutron experiments have reported that the both ordered phases
are associated with the identical wave vectors ${\bm
 Q}_0$=(0,0,1).~\cite{vilaume08} 
 First-principles calculations also predict the ${\bm Q}_0$ as the
 ordering vector from the calculations of ${\bm Q}$ dependent
 multipole susceptibility.~\cite{ikeda12}
 Interestingly, the de Haas-van Alphen (dHvA) measurements under pressure
 observe very little change in the frequencies from the HO phase to the
 ordinary AF magnetic phase,
 which indicate that the both phases have the very similar Fermi
 surfaces.~\cite{nakashima03, jo07, hassinger10}
 Fermi surfaces calculated with the local spin density
 approximation (LSDA) for the AF magnetic dipolar states well explain the dHvA
 frequencies,~\cite{yamagami00, elgazzar09, oppeneer10} and the energy
 bands are shown to be qualitatively unchanged for 
 the difference of the local multipoles, including the magnetic dipoles.~\cite{ikeda12}
  In this regard, electronic structure analysis also support the
 ${\bm Q}_0$ as the characteristic wave vector of the HO phase.
 Meanwhile, the ${\bm Q}$ vector characterizing HO phase
 transition is still open question,~\cite{mineev05,varma06,hanzawa07,schmidt10}
 and a direct experimental probe of the characteristic ${\bm Q}$ vector,
 for instance, by detecting the correspondent electronic/magnetic diffraction
 pattern is still important to clarify the HO state.

Possible signatures from the HO state have been reported in some
experimental studies.
The Ru-NQR and Si-NMR experiments have reported a broadening of the spectrum
below the transition temperature,~\cite{saitoh05,takagi07}
which indicates the presence of the internal fields induced with the HO.
 The elastic anomalies observed at T$_{\rm HO}$ can be related to the
 induced quadrupole moments accompanied by the phase
 transition.~\cite{luthi93, kuwahara97}
 These experimental observations, however, are considered to be
 secondary effects parasitic on the genuine order parameter of the HO
 state. Due to the extremely subtle changes, it is always difficult 
to conclude whether the secondary effects are intrinsic or extrinsic.
 Recent magnetic torque measurement reports broken in-plane 4-fold
 symmetry below T$_0$, suggesting the HO state has, so called, a nematic
 property of the electronic state.~\cite{okazaki11}
  This observation strongly confines symmetries in the HO phase,
 and therefore can be crucial clue to identify the genuine order
 parameter in the HO phase. Some recent experiments may have also
 been providing useful clues to confine possible symmetries of the HO
 states.~\cite{yanagisawa13, buhot14}

In order to understand these experimental observations, it is very
useful to clarify the global/site symmetry and the corresponding
secondary order parameters in high-rank MPO states.
Also, such classification would further stimulate experimental studies.
In this paper, we investigate the symmetry related properties of the
AF-MPO states of URu$_2$Si$_2$ by applying Shubnikov group theory, which
indicates full symmetry information concerning the electric/magnetic
degrees of freedom in the ordered states. Classification of the
AF-MPO states provides common ground for discussion of the hidden order
in URu$_2$Si$_2$.
For instance, we obtain knowledge regarding secondary order
parameters allowed to appear at each atomic site in several AF-MPO
states, which is crucial to experimentally identify the HO state.

  As discussed later, the charge distribution of AF-{\it magnetic} MPO states in
 URu$_2$Si$_2$ are clearly distinguished from the magnetic distribution
 in terms of symmetry, and the {\it magnetic} AF-$\Gamma^{-}$ MPO states
 always preserve the symmetry {of the charge distribution} higher than
 that of the corresponding electric AF-$\Gamma^{+}$ MPO states.
 On this point, occurrence of the AF-{\it magnetic} MPO states works better for
 concealing the ordered states by preserving the charge symmetries higher than
 ones of the electric MPO state.

%
%
 In Sec.\ \ref{Sec:MPO_SSG} symmetry of the MPO states
 in URu$_2$Si$_2$ are discussed based on Shubnikov group theory.
 After providing some definition regarding local multipoles defined on
 $U$ atoms in Sec.\ \ref{Sec:MPO_Local}, we discuss the global
 symmetry of the MPO states in Sec.\ \ref{Sec:MPO_Global}. 
 Features of secondary order parameters, of which the definition through
 this paper would be provided, in the AF-MPO states is
 discussed in Sec.\ \ref{Sec:ChargeSymmetry}.
 In Sec.\ \ref{Sec:MPO_SiteSym} the local site symmetries of Ru and Si
 sites are investigated to provide crucial information for indirect
 observation of the MPO states such as NQR/NMR experiment.
In Sec.\ \ref{Sec:calcSecondaryOP} we provide the computational
 analysis of A$_2^{-}$ and E$^{-}$ MPO states, in which the
 magnetic moments can be contained in terms of the symmetry, based on
 the first-principles approach to estimate
amount of the secondary order parameters in these MPO states.
It would be shown that the magnetic moments in the HO state are
 extremely small even if the symmetry allows the presence.
%
%
\section{Symmetric properties of AF-MPO states in URu$_2$Si$_2$}
\label{Sec:MPO_SSG}
\subsection{Multipole theories of URu$_2$Si$_2$}
\label{Sec:MPO_Local}
\begin{figure}[t]
\begin{center}
\includegraphics[width=1.0\linewidth]{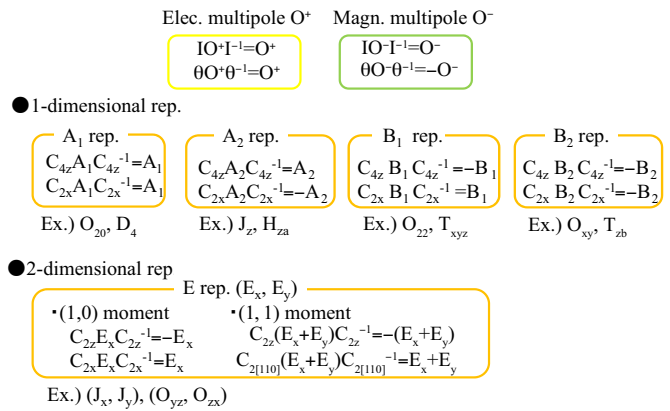}
\caption{ Local transformation property of multipoles in a $D_{4}$ point
 symmetry site. The multipole operators are
 represented as the symbol of each irreducible representation, whose
 examples are shown below each box. The symbols of multipole operators are defined
 following TABLE S1 in the supplementary information of Ref. \onlinecite{ikeda12}.}
\label{Fig:SSG3}
 \end{center}
\end{figure}

 The multipole moments of URu$_2$Si$_2$ are classified according to
 IREP of the D$_{4h}$ point group to which the U sites belong,
 i.e. $\Gamma$=$A_{1g}$, $A_{2g}$, $B_{1g}$, $B_{2g}$, or $E_{g}$, 
 assuming the presence of the space inversion symmetry in the HO phase.
 We would omit the index $g$ hereafter.
  The multipole moments are further divided by parity for the time
 reversal operation, and the multipole which have the even/odd parity
 for the time reversal symmetry would be indicated by the index $+$/$-$,
 such as $\Gamma^{+}$ or $\Gamma^{-}$.
 For instance, the magnetic dipole $J_z$ belongs to $A_2^-$ multipole.
  Note that the multipole with E-IREP, proposed as primary
 order parameter in our previous study,~\cite{ikeda12} has
 two-dimensional IREP, and the symmetry depends on the linear
 combination of (E$_{x}$, E$_{y}$), see Fig.\ \ref{Fig:SSG3}. We take,
 through this paper, the two dimensional components (E$_{x}$, E$_{y}$)=(1,1), which is
 consistent with the nematic properties observed experimentally.~\cite{okazaki11, tonegawa12, kambe13}

 We focus on, through this paper, the MPOs which are characterized
by the commensurate wave vector ${\bm Q}$=${\bm Q}_0$=(0,0,1), which is
considered to be a plausible possibility from the previous experimental and
theoretical works.~\cite{vilaume08,ikeda12} 
As a matter of fact, the rotation symmetries which distinguish the type
of multipoles as in Fig.\ \ref{Fig:SSG3} can be present only with the
translation symmetries preserved in the order characterized by, except
${\bm Q}$=${\bm 0}$, the minimal commensurate wave vector, and one of
the main arguments in this paper, suggesting the high symmetry of
the AF-MPO states, is therefore not held with other finite ${\bm Q}$.
The fact that the HO phase is characterized by the wave vector ${\bm
Q}_0=(0,0,1)$ leads to a simple consequence that the local order
parameters aligned on U sites with simple type-I AF structure, in which the
multipole moments on the body-center uranium sites are {\it
opposite} from ones on the other uranium sites,  in the HO phase. 
Here, {\it opposite} multipole moment is defined as the
multipole moment of which the expectation value has opposite sign.
 The second order phase transition at T$_{\rm HO}$ restrict the symmetry
 of the HO phase to a subgroup of the paramagnetic space group above T$_{\rm HO}$.
 The Landau phenomenological analysis further indicates that the
 bi-critical point of the paramagnetic, the HO, and the AF magnetic (AFM) phases on the P-T phase diagram~\cite{motoyama03, amato04, motoyama08, hassinger08, butch10} is
 explained from the fact that the symmetry of the HO phase is different from that
 of the ordinary AFM phase.~\cite{mineev05, kiss05}
  These facts are consistent that the HO state is characterized by type-I AF structure of a
 multipole moment with symmetry {\it different} from that of the
 ordinary AFM state, and we hereafter consider symmetry in type-I AF MPO
 states in URu$_2$Si$_2$.

\subsection{Shubnikov space groups of AF-MPO states in URu$_2$Si$_2$}
\label{Sec:MPO_Global}
Shubnikov group theory can provide common ground for characterizing the
electric/magnetic symmetries in the paramagnetic, the AFM, and the
  HO phases of URu$_2$Si$_2$ in a unified manner.
 In general, the Shubnikov space group (SSG) $\mathcal{M}$ are classified in
 relation to the ordinary space group (OSG) $\mathcal{G}$ into two broad
 categories depending on the presence or absence of anti-unitary part as
 follows:
\begin{eqnarray}
\mathcal{M} = \mathcal{G},
\label{Eq:TypeISSG} \\
\mathcal{M} = \mathcal{G} + \theta\{ R\mid{\bm \tau}\} \mathcal{G}
\label{Eq:SSGroup},
\end{eqnarray}
where $\theta$ is time reversal operator and $\{ R\mid{\bm \tau}\}$ is a
space group element with the rotation operator $R$ accompanied by the
extra translation ${\bm \tau}$. Note that the $\{ R\mid{\bm \tau}\}$ is
not necessary to be an element of the unitary group $\mathcal{G}$ in
eq. (\ref{Eq:SSGroup}).
  For instance, the symmetry of paramagnetic phase is characterized by
 a paramagnetic space group $\mathcal{G}_{\rm PM}\otimes T$,~\cite{bradley1972_book} where the
 $\mathcal{G}_{\rm PM}$ is the OSG of the crystal
 structure, which is  $I4/mmm$ (No. 139) in case of URu$_2$Si$_2$, and $T$=$\{E$,$\theta \}$. 
 The paramagnetic space group corresponds to the SSG of
 eq. (\ref{Eq:SSGroup}) with $\{ R\mid{\bm \tau}\}=\{E\mid{\bf 0}\}$.
   As we will see in the case of the AF-MPO states of URu$_2$Si$_2$, 
  the global symmetry of MPO state is determined by the local
  transformation property of the order parameter for the rotation operators
  of space group (See Fig.\ \ref{Fig:SSG3}).

 The SSG of Eq. (\ref{Eq:TypeISSG}), which is characterized by
 absence of the anti-unitary part, is called Fedrov group.~\cite{bradley1972_book} 
 For instance, the 3${\bf q}$ type magnetic multipolar state of NpO$_2$
 belongs to Fedrov group of eq. (\ref{Eq:TypeISSG}) since any
 anti-unitary operations do not preserve the non-collinear alignment of
 the magnetic multipoles.
  The SSG of HO phase in URu$_2$Si$_2$ belongs to a
 subgroup of the paramagnetic space group from the discussion of Sec.\
 \ref{Sec:MPO_Local}, and the possible AF-MPO states of URu$_2$Si$_2$ belong to
 the SSG of eq. (\ref{Eq:SSGroup}). 
This is because the time reversal operation, which keeps electric
 multipoles and flips magnetic multipoles, preserve the global symmetry
 in case of electric MPO and with an extra translation to the body center sites, ${\bm \tau}_{\rm AF}$, in magnetic MPO. 

  As discussed in sec.~\ref{Sec:ChargeSymmetry}, the different
 categories of the SSG for the AF-MPO states in
 NpO$_2$ and in URu$_2$Si$_2$ clearly distinguish the property of
 secondary order parameters allowed in these ordered states.

 The electric and magnetic MPO states in URu$_2$Si$_2$ are thus clearly
 distinguished through the anti-unitary part of SSG as
 following:
\begin{eqnarray}
\mathcal{M}_{\Gamma^{\pm}} = \mathcal{G}_{\Gamma} + \theta\{ E \mid {\bm \tau^{\pm}} \}\mathcal{G}_{\Gamma}\ ,
\label{Eq:SSG_MPO}
\end{eqnarray}
 where ${\bm \tau^{+}} = {\bm 0}$ for electric MPO and ${\bm \tau^{-}} =
 {\bm \tau}_{\rm AF} $ for magnetic MPO, respectively.
 The space group $\mathcal{G}_{\Gamma}$ corresponds to $P4/mmm$
 (No.123), $P4/mnc$ (No.128), $P4_{2}/mmc$ (No.131), $P4_{2}/mnm$
 (No.136), and $Cmce$ (No.64) for $\Gamma =A_{1}$, $A_{2}$, $B_{1}$,
 $B_{2}$, and $E$ multipolar states, respectively.
  Note that the space groups $\mathcal{G}_{\Gamma}$ except the
 $\mathcal{G}_{A_1}$ belong to nonsymmorphic space groups with extra
 translation ${\bm \tau_{AF}}$.
 Since the electric MPOs do not break magnetic symmetry, the full
 symmetry can be described by the OSG $\mathcal{G}_{\Gamma}$
 within electric degrees of freedom, and the electric symmetries
 have been investigated in earlier work.~\cite{harima10}
  The magnetic symmetries correspond to the MPO states of
 magnetic multipoles which belong to one-dimensional IREPs are also explored in
 the magnetic neutron study.~\cite{khalyavin14}
 Our present work has developed the symmetry analysis so as to
 describe possible electric/magnetic multipolar
 states in the same framework, clarifying the structure of global symmetries
 of the electric and magnetic MPO states.

\subsection{Secondary order parameters in magnetic MPO states}
\label{Sec:ChargeSymmetry}

   Order parameters characterize symmetry breaking from a paramagnetic
  state to ordered states. In the ordered states, there often appear
  additional changes induced by the presence of the primary order
  parameters, such as structural distortion in magnetic orderings.
  Such secondarily induced quantities are compatible with symmetry in
  the ordered states. Considering the global/site symmetry in the
  ordered states, we can classify the presence of the induced secondary
  order parameters. Knowledge of such secondary order parameters is
  crucially important in high-rank MPO states, since it is very
  difficult to detect high-rank multipole moments experimentally.
  The SSG classification of the MPO states is available for
 identification of all of the secondary order parameters in the MPO states.

  Electric multipole moments induced as the secondary order
 parameters often provides crucial information to identify
 the magnetic MPO, since the reduced charge symmetry can
 be detected through change in the electronic diffraction pattern. 
 Furthermore, detection of electric quadrupole moments are
 generally easier than detection of the higher rank magnetic
 multipoles by the experiment such as resonant X-ray or ultrasonic sound wave.
 The symmetry of the electric component in magnetic MPO is, in
 general, derived from the SSG.
 In case of the magnetic ordered states which belongs to Fedrov group,
 the OSG of charge distribution is indicated by
 $\mathcal{G}$, and the symmetry belongs to the same one with the
 magnetic distribution as seen in Eq. (\ref{Eq:TypeISSG}).
 For instance, because the 3{\bf q} AF magnetic structure in
 NpO$_2$ belongs to Fedrov group, the OSG of the MPO ground
 states~\cite{nikolaev03} already provides full symmetry of the magnetic MPO states of NpO$_2$ from eq. (\ref{Eq:TypeISSG}).
  In this case, the electric multipoles which belong to the same
 symmetry with the primary magnetic multipole is induced as the
 secondary order parameter.
 Actually, resonant X-ray and NMR experiments concluded the magnetic
 multipolar state of NpO$_2$ by detecting the
 {\it electric} quadrupole order induced with the primary higher-rank
 magnetic MPO,~\cite{paixao02, tokunaga05, tokunaga06} considering
 the experimental observations of the breaking of time reversal
 symmetry.~\cite{erdos80, kopmann98}
\begin{widetext}
  \begin{center}
  \begin{table}[t]
   \begin{tabular}{|cc|} \hline
   \multicolumn{2}{|c|}{Paramagnetic phase} \\
   $\mathcal{M}$     & $\mathcal{G}_{C}$ \\ \hline 
 $I4/mmm1'$ & $I4/mmm$ \\ \hline
   \end{tabular}
   \begin{tabular}{cc} \hline
   \begin{tabular}{ccccc} \hline
   \multicolumn{5}{c}{AF-Electric MPO phase} \\
    IREP      & $\mathcal{M}$ & $\mathcal{G}_{C}$ & Time reversal & 4-fold rotation \\ \hline
 A$_{1}^{+}$  & $P4/mmm1'$ & $P4/mmm$ & Preserved &  Preserved \\
 A$_{2}^{+}$  & $P4/mnc1'$ & $P4/mnc$ & Preserved &  Preserved \\
 B$_{1}^{+}$  & $P4_{2}/mmc1'$ & $P4_{2}/mmc$ & Preserved &  Preserved \\
 B$_{2}^{+}$  & $P4_{2}/mnm1'$ & $P4_{2}/mnm$ & Preserved &  Preserved \\
 E$^{+}$      & $Cmce1'$ & $Cmce$ & Preserved & Broken   \\ \hline 
   \end{tabular}
   \begin{tabular}{ccccc} \hline
   \multicolumn{5}{c}{AF-Magnetic MPO phase} \\
    IREP    & $\mathcal{M}$ & $\mathcal{G}_{C}$ & Time reversal & 4-fold rotation \\ \hline 
 A$_{1}^{-}$  & $P_{I}4/mmm$ & $I4/mmm$  & Broken    & Preserved    \\
 A$_{2}^{-}$  & $P_{I}4/mnc$ & $I4/mmm$  & Broken    & Preserved  \\
 B$_{1}^{-}$  & $P_{I}4_{2}/mmc$ & $I4/mmm$ & Broken     & Preserved  \\
 B$_{2}^{-}$  & $P_{I}4_{2}/mnm$ & $I4/mmm$ & Broken     & Preserved  \\
 E$^{-}$      & $C_{A}mce$  & $Fmmm$ & Broken & Broken \\ \hline
   \end{tabular}
   \end{tabular}
    \caption{SSG of the magnetic symmetries, $\mathcal{M}$, in the AF-MPO states of
   URu$_2$Si$_2$ and the CSG for the charge distribution, $\mathcal{G}_{C}$. 
    The notation of SSG follows Ref.\ \onlinecite{bradley1972_book}. 
   The presence of time reversal and four fold rotational symmetries
   are also indicated.}
\label{tab:GlobalSym}
  \end{table}
  \end{center}
\end{widetext}

\begin{figure}[b]
\begin{center}
\includegraphics[width=1.0\linewidth]{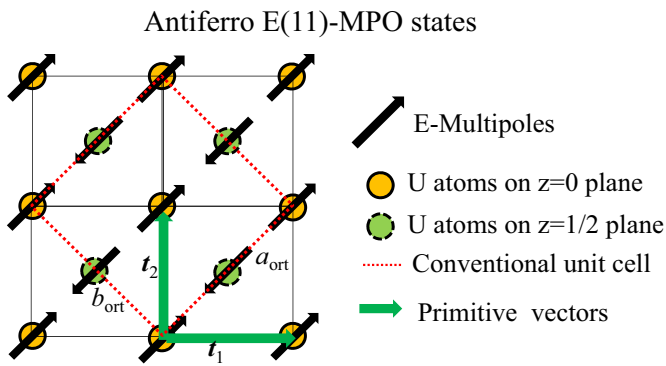}
\caption{Schematic picture of the relation between the primitive and
 the conventional unit cell in AF-E MPO induced with the
 two dimensional components ($E_x$, $E_y$)=(1,1).}

\label{Fig:EminusMPO}
 \end{center}
\end{figure}

 In case of the magnetic states which belong to Eq. (\ref{Eq:SSGroup}),
 the symmetry of the charge distribution is indicated by the OSG
 $\mathcal{G}+\{ R\mid{\bm \tau}\} \mathcal{G}$ where the symbols
 are already defined in Eq. (\ref{Eq:SSGroup}), since the time
 reversal operation does not affect to the charge component.
 We will refer to the OSG of charge distribution as charge space group
 (CSG) $\mathcal{G}_{C}$ hereafter.
  From above discussion, the symmetry of charge
  distribution in each AF-MPO state of URu$_2$Si$_2$ is identified.
 The SSG of the magnetic symmetry in the AF-MPO states and the corresponding CSG
  are listed in Table \ref{tab:GlobalSym}.
 It is found from the table that the magnetic MPO characterized by $A_{1}^{-}$,
 $A_{2}^{-}$, $B_{1}^{-}$, and $B_{2}^{-}$-MPOs, which
 belong to one dimensional IREPs, have the same CSG with that of
 paramagnetic states, i.e. $\mathcal{G}_{C}=I4/mmm$. This fact
 leads to a conclusion that these AF-MPO states do not induce any charge
 deformation which break the symmetry of the paramagnetic states.
  In other words, identification of these MPO  states is possible
 only by detecting the symmetry breaking caused by the magnetic distribution.

 The AF E$^{\pm}$ multipolar states break the four-fold
 rotational symmetry as shown in the table \ref{tab:GlobalSym}.
 When the multipole moments are induced with the two dimensional components
 preserving the symmetry higher, such as
 (E$_x$,E$_y$)=(1,1), the $E^{\pm}$ MPO
 states have orthorhombic conventional cell, yet the primitive unit cell
 is the same with that of the other multipolar states, see Fig. \ref{Fig:EminusMPO}.
 Operation of two fold rotation along the tetragonal axis, $C_{2z}$,
 transform the E multipoles to the {\it opposite}, and
 the unitary operator $\{C_{2z}\mid {\bm \tau}_{\rm AF}\}$
 therefore preserve the AF-$E$ MPO states equivalent.
  Okazaki et al. refer the fourfold symmetry breaking as
 {\it electronic nematic}.~\cite{okazaki11}
 The AF-E$^{-}$ MPO state, interestingly, permit an electronic
 deformation to the ferroic B$_{2}^{+}$ MPO, which contain the {\it
 ferroic} $O_{xy}$ quadrupole order, since the charge distribution in
 the AF-E$^{-}$ MPO states belongs to the space group $Fmmm$ (No. 69). 
 This electronic deformation is symmetrically
 equivalent to that reported in the iron-based superconductor
 BaFe$_2$As$_2$.~\cite{yoshizawa12}
 The recent highly accurate X-ray diffraction measurement using
 high-resolution synchrotron X-ray has reported the
 observation of the reduction of electric symmetry in the HO
 phase of URu$_2$Si$_2$.~\cite{tonegawa14}

\subsection{Atomic site symmetries in AF-MPO states}
\label{Sec:MPO_SiteSym}
 The local symmetries of Ru and Si atomic sites in the HO phase are important to
 identify the ordered states through the observation of the
 secondarily induced electric/magnetic distribution on the atomic
 sites.~\cite{saitoh05, takagi07} 
 The local magnetic/electric symmetry of MPO states are also derived from
 the SSG in the ordered states.

 Table \ref{tab:LocalSym} shows the site symmetries of all atomic sites
 in each AF-MPO state of URu$_2$Si$_2$.
 The ordinary point groups for local charge distribution at each
 atomic site are also found from this table by omitting $1'$ and the primes (') from the symbols.
 Since the multipole moments in URu$_2$Si$_2$ is originated from U-$f$
 electrons, the local symmetries of U sites are equivalent to that of
 the multipole moments. 
 The magnetic dipoles along the tetragonal axis can be induced on the U
 and Si atomic sites in the AF-A$_2^{-}$ MPO states, to which the
 ordinary AFM states belong, and on the Ru sites in the AF-B$_2^{-}$ MPO
 states. Moreover, {\it in-plane} magnetic dipole moments are allowed in
 the all of atomic sites in the AF-E$^{-}$ MPO states. 
 In these AF-MPO states, the secondarily induced dipole moments can be
 detected by experimental observation by NQR/NMR experiments although
 the induced signals may be extremely small.~\cite{kambe13, mito13}

 On the other hand, no magnetic dipole moment is induced at any atomic
 sites in AF-A$_{1}^{-}$ and AF-B$_{1}^-$ MPO states. 
 Since these AF-MPO states neither induce any charge deformations which
 break the symmetry of paramagnetic states as discussed in Sec.\
 \ref{Sec:ChargeSymmetry} and as also found in table I and II,
 identification of these ordered states requires, for experiments,
 direct detection of the higher rank magnetic multipoles such as
 magnetic octupole and magnetic triakontadipole in case of the B$_{1}^-$
 and A$_{1}^{-}$ MPO, respectively (see, e.g., TABLE I in Ref.\
 \onlinecite{thalmeier11} or TABLE S1 of supplementary information of
 Ref. \onlinecite{ikeda12}).

\begin{widetext}
  \begin{center}
  \begin{table}[htb]
  \begin{tabular}{|ccc|} \hline
 \multicolumn{3}{|c|}{Paramagnetic phase} \\
 Atom& Atomic sites & Local sym.  \\ \hline
 U   & 1 & $(2a)\ 4/mmm1'$  \\
 Ru  & 1 & $(4d)\ \bar{4}m21'$  \\
 Si  & 1 & $(4e)\ 4mm1'$   \\ \hline
  \end{tabular}
\begin{tabular}{cc}
  \begin{tabular}{ccccc} \hline
 \multicolumn{5}{c}{AF-Electric MPO phase} \\
              & Atom & Atomic sites & Local sym.  & Mag. moments \\ \hline
              & U  & 2 & $(1a)\ 4/mmm1' \oplus (1d)\ 4/mmm1'$ & None \\
 A$_{1}^{+}$  & Ru & 1 & $(4i)\ 2mm.1'$                  & None \\
              & Si & 2 & $(2g)\ 4mm1'\oplus (2h)\ 4mm1'$ & None \\
              & U  & 1 & $(2a)\ 4/m..1'$  & None \\
 A$_{2}^{+}$  & Ru & 1 & $(4d)\ 2.221'$ & None \\
              & Si & 1 & $(4e)\ 4..1'$  & None \\
              & U  & 1 & $(2c)\ mmm.1'$ & None \\
 B$_{1}^{+}$  & Ru & 2 & $(2e)\ \bar{4}m21'\oplus (2f)\ \bar{4}m21'$ & None \\
              & Si & 1 & $(4i)\ 2mm.1'$ & None \\
              & U  & 1 & $(2a)\ m.mm1'$ & None \\
 B$_{2}^{+}$  & Ru & 1 & $(4d)\ \bar{4}..1'$ & None \\
              & Si & 1 & $(4e)\ 2.mm1'$ & None \\
              & U  & 1 & $(4a)\ 2/m..1'$ & None \\
 E$^{+}$      & Ru & 1 & $(8e)\ .2.1'$   & None \\
              & Si & 1 & $(8f)\ m..1'$ & None \\ \hline
  \end{tabular}
  \begin{tabular}{ccccc} \hline
 \multicolumn{5}{c}{AF-Magnetic MPO phase} \\
              &Atom & Atomic sites & Local sym. & Mag. moments \\ \hline
              & U  & 1 & $(2a)\ 4/mmm$       & None \\
 A$_{1}^{-}$  & Ru & 1 & $(4d)\ \bar{4'}m2'$ & None \\
              & Si & 1 & $(4e)\ 4mm$ & None \\
              & U  & 1 & $(2a)\ 4/mm'm'$ & $M_{z}$ \\
 A$_{2}^{-}$  & Ru & 1 & $(4d)\ \bar{4'}m'2$ & None \\
              & Si & 1 & $(4e)\ 4m'm'$   & $M_{z}$ \\
              & U  & 1 & $(2a)\ 4'/mmm'$ & None  \\
 B$_{1}^{-}$  & Ru & 1 & $(4d)\ \bar{4}m2$ & None \\
              & Si & 1 & $(4e)\ 4'mm'$ & None \\
              & U  & 1 & $(2a)\ 4'/mm'm$ &  None \\
 B$_{2}^{-}$  & Ru & 1 & $(4d)\ \bar{4}m'2'$ & $M_{z}$ \\
              & Si & 1 & $(4e)\ 4'm'm$ & None \\
              & U  & 1 & $(4a)\ mm'm'$  & $M_{x{\rm (Ortho)}}$ \\
 E$^{-}$      & Ru & 1 & $(8f)\ 2'22'$  & $M_{y{\rm (Ortho)}}$ \\
              & Si & 1 & $(8i)\ mm'2'$  & $M_{x{\rm (Ortho)}}$ \\ \hline
  \end{tabular}
\end{tabular}
    \caption{Local symmetries of atomic sites in the AF-MPO states. The
   third column shows number of the non-equivalent atomic sites.
   The notation of site symmetry basically follows
   Ref. \onlinecite{internationaltables2002}, adding $1'$ for the pure
   time reversal operator or prime (') on the
   symbols corresponding to operators combined with time reversal.
   Wyckoff letters for CSG are adapted to specify the atomic positions,
   see also Table \ref{tab:GlobalSym}.
    The magnetic moments of $E^{-}$-MPO, M$_{\mu ({\rm Ortho})}$
   ($\mu$=x, y), are labeled for the {\it orthorombic} axis (see Fig.\
   \ref{Fig:EminusMPO}), which have relation such as M$_{x ({\rm
   Ortho})}$=$\frac{1}{\sqrt{2}}(M_x+M_y)$. }
\label{tab:LocalSym}
  \end{table}
  \end{center}
\end{widetext}

\section{Computational investigation of secondary order parameters}
\label{Sec:calcSecondaryOP}
 Recent magnetic torque measurement and high-resolution synchrotron X-ray
 diffraction measurement have reported observation of breaking four-fold rotation
 symmetry in the HO phase.~\cite{okazaki11, tonegawa14}
 Within the multipole theory, these observations indicate that the order
 parameters are E-type multipoles.  
  The rather low charge symmetry of AF-E$^{+}$ MPO, which are shown in Table
 \ref{tab:GlobalSym} and \ref{tab:LocalSym}, not only for rotation symmetry
 but also for translation symmetry is likely to be difficult to
 conceal the order parameters from experimental observations, but it is
 still a possible candidate in terms of the breaking four-fold symmetry including the observation of
 split of a X-ray diffraction spot~\cite{tonegawa14} and, possibly, lack of magnetic moments.

 Our previous theoretical works have predicted the magnetic E$^-$ MPO state in the HO
 phase.~\cite{ikeda12}
 In terms of symmetry, the E$^-$-MPO states contain in-plane magnetic moments
 as the member of multipoles, and {\it why experiments have not detected the
 in-plane dipole moments so far} is a big issue to be resolved if the 
 AF-E$^-$-MPO is the ordered state in the HO phase.
 To investigate the problem, we have performed the first-principles
 electronic structure calculations to quantitatively estimate the
 magnetic moments which can be present in the AF-MPO states. 
 Since the electric/magnetic degree of freedom to describe the
multipole moments are included in the local LDA+$U$ potentials, the AF-MPO states
can be calculated with the LDA+$U$ calculations properly considering the global
electric/magnetic symmetries.~\cite{mtsuzuki10,mtsuzuki13}
The detailed calculation method for the multipole ordered states based
on the LDA+{\it U} method is described in Ref. \onlinecite{mtsuzuki13}.
We used the exchange-correlation functional of Gunnarsson and Lundqvist
for the LDA potential.~\cite{gunnarsson76}
The double-counting term has been chosen as in the around mean field,
leaving out the spin dependency of the Hund's coupling part to adapt it
for the nonmagnetic LDA part as in the Ref. \onlinecite{mtsuzuki13}.
 Earlier theoretical work has estimated the appropriate range of $U$
 values of URu$_2$Si$_2$ as less than 1 eV.~\cite{cricchio09} We
 therefore performed the calculations by changing the $U$ values from 0
 to 1.0eV with the Hund's coupling parameter $J$=0.1eV.
 The ordered states are calculated under the restriction of SSG
 for each AF-MPO state.

 The multipole moments can be described through the radial function of
 the local density matrix, calculated as follwing 
\begin{eqnarray}
\rho^{\tau\ell}_{\gamma\gamma'}(r^{\tau})= \frac{1}{N}
 \sum_{\boldsymbol{k}b}\langle \tau\ell\gamma | \boldsymbol{k}b \rangle
 \ \langle \boldsymbol{k}b | \tau\ell\gamma'\rangle (r^{\tau}) ,
\end{eqnarray}
where $\tau$ and $\ell$ denote the atoms and angular momenta of the
orbitals, respectively. $\gamma$ ($\gamma'$) is an index related to an orbital $m$
($m^{\prime}$) and a spin $s$ ($s'$). 
$\boldsymbol{k}$ and $b$ denote wave vector and band index,
respectively.
$N$ is the number of $\boldsymbol{k}$ points
and $r^{\tau}$ the radial component of the position vector $\boldsymbol{r}^{\tau}$ measured from atom $\tau$.
The local multipole moment $O^{\tau}$ on a atom is calculated with the local basis set \{$|\tau\ell\gamma \rangle$\} inside the MT sphere, following the expression~\cite{mtsuzuki13}
\begin{eqnarray}
O^{\tau} = \!\!\!\sum_{\ell}\sum_{\gamma} \sum_{\gamma_1\gamma_2} \int
 \! dr^{\tau} \{r^{\tau}\}^{2}\rho^{\tau\ell}_{\gamma\gamma_1}(r^{\tau})
 O^{\tau\ell}_{\gamma_1\gamma_2}\rho^{\tau\ell}_{\gamma_2\gamma}(r^{\tau}) ,
\label{Eq:MPExpect}
\end{eqnarray}
where $O^{\tau\ell}_{\gamma\gamma'}$ denote the matrix elements of
the multipole operator $\hat{O}^{\tau\ell}$, whose explicit expressions
up to rank-5 are listed in the supplementary information of our previous work.~\cite{ikeda12}
\begin{figure}[t]
\begin{center}
\includegraphics[width=0.7\linewidth]{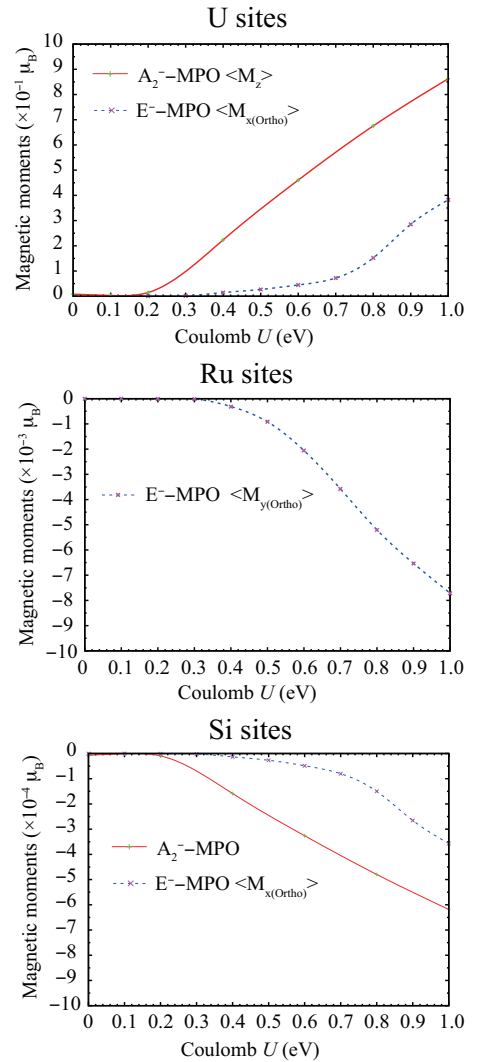}
\caption{Calculated magnetic moments symmetrically arrowed to appear on
 each atomic site in the AF-A$_{2}^{-}$ and the AF-E$^{-}$
 states. Hund's coupling $J$ is chosen as 0.1eV in the LDA+$U$
 method.}
\label{Fig:MagMoments}
 \end{center}
\end{figure}
 Figure \ref{Fig:MagMoments} plots magnetic moments on each atomic site in the A$_2^{-}$ and E$^{-}$ MPO states of URu$_2$Si$_2$ applying the $U$ values from 0 to 1eV. 
 In the symmetry view point, the magnetic moments can be present on U
 and Si sites in the A$_2^{-}$-MPO, on Ru sites in B$_{2}^{-}$-MPO, and
 on all atomic sites in E$^{-}$-MPO states. Meanwhile, our calculations
 show no finite values for $<M_z>$ moments on Ru atoms in
 B$_{2}^{-}$-MPO states for the calculated $U$ region.
  This result shows that the B$_2^{-}$ multipoles on U sites, which
 contain no dipole moment as the constituent, hardly couple with
 the magnetic dipoles on the Ru sites.
  In the A$_2^{-}$-MPO states corresponding to AFM phase, the magnetic
 moments along the tetragonal axis linearly increase with the increased
 $U$ values above 0.2eV. Amount of the magnetic moments are
 $\sim$10$^{-1} \mu_B$, which is consistent with experimentally observed
 magnetic moments in the AFM phase under pressure,
 i.e. 0.1$\sim$0.4$\mu_B$ for the small range of $U$ below 0.6eV.
 On the other hand, the shift of in-plane magnetic moments of
 E$^{-}$-MPO states is suppressed below 0.7eV and
 develop linearly above 0.7eV, in which the magnetic moments are at most one order smaller than ones of
 the A$_2^{-}$ states on U atoms. 
 This result shows that the paramagnetic electronic structure around the Fermi level
 has tiny response for the in-plane magnetic moments, which is
 consistent with our previous work.~\cite{ikeda12}
\begin{figure}[t]
\begin{center}
\includegraphics[width=1.0\linewidth]{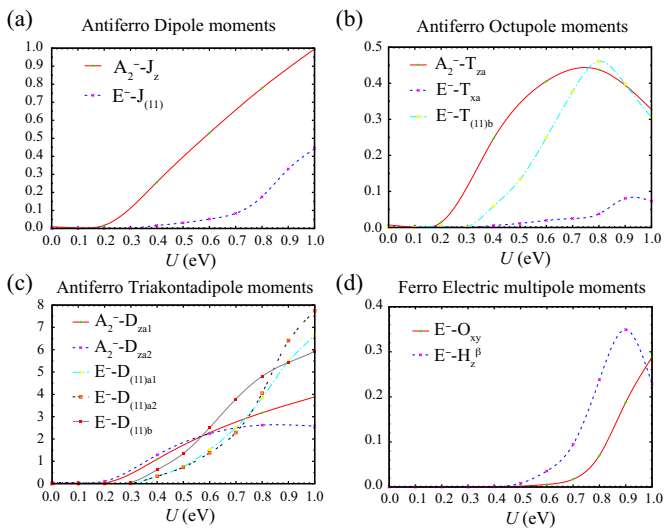}
\caption{Absolute values of calculated (a) dipole, (b) octupole, (c)
 triakontadipole moments induced on uranium sites in the AF-A$_{2}^{-}$
 and AF-E$^{-}$ states. The ferroic electric multipolar moments in the
 E$^{-}$ multipolar states are also plotted in (d). The operator
 expression of multipole
 moments are defined for the tetragonal axis, and symbols such as
 $A_{(11)}$ indicate the multipole moments with the components
 ($A_x$, $A_y$)=(1, 1) in two dimensional IREP.,
 i.e. $A_{(11)\alpha}\equiv \frac{1}{\sqrt{2}}(A_{x\alpha}+A_{y\alpha})$.}
\label{Fig:MPMoments}
 \end{center}
\end{figure}

The magnetic moments secondarily induced by the MPO at U sites are also
plotted in the Fig.\ \ref{Fig:MagMoments}.
This figure shows that the LDA+$U$ calculation for the E$^{-}$-MPO states produces the magnetic moments $\sim$
10$^{-2}$ $\mu_B$ at U sites,  $\sim$ 10$^{-4}$ $\mu_B$ at Ru sites and
$\sim$ 10$^{-5}$ $\mu_B$ at Si sites for the expected $U$ values
discussed above.
 There is an obvious trend to leave the induced moments small, as compared with those of AFM states in the appropriate $U$ range. Although the magnitude of the induced moment on U site may be detectable, it is likely that the effect of electron correlations beyond the Hartree-Fock level in the LDA+$U$ method makes this value still one-order smaller, as such effect is important to suppress the in-plane magnetic response in the uniform susceptibility.

  Figure\ \ref{Fig:MPMoments} shows that, while the development of the
  in-plane magnetic dipole moments of E$^-$-MPO for the $U$ parameter 
  is considerably smaller than that of A$_2^-$-MPO, the higher-rank multipole 
 moments have comparable contribution of octupole moments and larger
 contribution of triakontadipole moments compared to those of A$_2^-$-MPO.
We have also confirmed the electric multipolar moments accompanied by the
E$^{-}$-MPO is also very little in the AF-E$^{-}$ MPO states as shown in
Fig.\ \ref{Fig:MPMoments} (d). 
Although possible lattice distortions of $a_{\rm ortho}\ne b_{\rm
  ortho}$, which is not taken into account in the current
results, can slightly enhance the ferroic quadrupole moments, we find
the lattice distortion in E$^-$-MPO state is extremely small due to the
very weak coupling between the quadrupole moments and the lattice system. 
Nevertheless, the experimental techniques with great sensitivity for
the electronic modification such as NQR/NMR ~\cite{saitoh05, takagi07}
or ultrasonic sound wave experiments~\cite{yanagisawa13, kuwahara97}
may have captured the subtle trace from the reduction of charge
symmetry.
 These results indicate that the magnetic E$^-$-MPO states obtain the
 energy gain not from the dipole moment but from the higher-rank
 magnetic multipole moments in the small range of $U$.

%
%
\section{Summary}
We have investigated the character of the full symmetry in the
AF-MPO states as promising candidates of the hidden-order state in
URu$_2$Si$_2$. 
On the basis of Shubnikov group theory, the electric/magnetic symmetry
in the AF-MPO states has been explicitly identified, and the global and
atomic local symmetry have been classified.
The AF-MPO states which belong to one dimensional IREPs induce no
electric charge deformation, and $A_{1}^{-}$ and $B_{1}^{-}$ MPO states
do not allow any induced dipole moments at atomic sites.
Although AF-$E^{-}$ MPO states can induce in-plane AF dipole and
ferroic $B_{2}^{+}$ multipoles as the secondary order parameter, our
LDA+$U$ calculations revealed that the magnitude of the secondary
order parameters is extremely small in the appropriate interaction parameter range. 
Further experimental trials to detect possible traces from HO states are
strongly encouraged to identify the order parameters. 

%

\acknowledgments
{ We thank S.\ Kambe, Y. Tokunaga, and T. Takimoto for valuable discussions. This
work has been supported by JSPS KAKENHI Grant Number 24540369. }

\bibliographystyle{revtex}
\bibliography{URu2Si2}

\end{document}